\documentclass[aps,prc,twocolumn,groupedaddress,showpacs,eqsecnum]{revtex4}
\usepackage{graphicx}% Include figure files
\usepackage{dcolumn}% Align table columns on decimal point
\usepackage{bm}% bold math

\newcommand{\ling}{{$^{8}$Li($n$,$\gamma$)$^{9}$Li}}
\newcommand{\bpg}{{$^{8}$B($p$,$\gamma$)$^{9}$C}}
\newcommand{\lbo}{$\lambda_{\rm{bound}}$}
\newcommand{\lsc}{$\lambda_{\rm{scatt}}$}
\newcommand{\lbocent}{$\lambda_{\rm{bound}}^{\rm{central}}$}
\newcommand{\lbols}{$\lambda_{\rm{bound}}^{\rm{LS}}$}

\begin{document}

\title{
Low-Energy Direct Capture in the
$^{8}$Li($\bm n$,$\bm\gamma$)$^{9}$Li 
and
$^{8}$B($\bm p$,$\bm\gamma$)$^{9}$C 
Reactions
}

\author{P.\ Mohr}
\email[E-mail: ]{mohr@ikp.tu-darmstadt.de}
\affiliation{
Institut f\"ur Kernphysik, Technische Universit\"at Darmstadt,
Schlossgartenstra{\ss}e 9, D-64289 Darmstadt, Germany
}

\date{\today}

\begin{abstract}
The cross sections of the \ling\ and \bpg\ capture reactions have been
analyzed using the direct capture model. At low energies which is the
astrophysically relevant region the capture process is dominated by
$E1$ transitions from incoming $s$-waves to bound $p$-states. The
cross sections of both mirror reactions can be described
simultaneously with consistent potential parameters, whereas previous
calculations have overestimated the capture cross sections
significantly. However, the parameters of the potential have to be
chosen very carefully because the calculated cross section of the
\ling\ reaction depends sensitively on the potential strength.
\end{abstract}

% 26.30.+k  Nucleosynthesis in novae, supernovae and other explosive
%           environments
% 26.35.+c  Big Bang nucleosynthesis
%           (see also 98.80.F Origin and formation of the elements)
% 25.40.-h  Nucleon-induced reactions
% 25.40.Lw  Radiative capture

\maketitle

\section{\label{sec:intro}Introduction}
Nucleon capture in the \ling\ and \bpg\ mirror reactions has attracted
much attention in the recent years. The low-energy behavior of both
reactions is of astrophysical relevance. Nucleosynthesis of light
nuclei is hindered by the gaps at $A = 5$ and $A = 8$ where no stable
nuclei exist. However, these gaps may be bridged by reactions involving
the unstable $A = 8$ nuclei $^8$Li ($T_{1/2} = 840$\,ms) and $^8$B
($T_{1/2} = 770$\,ms).

The \ling\ reaction is important in inhomogeneous big bang models.
Here the \ling\ reaction \cite{Koba03} is in competition with the
$^8$Li($\alpha$,n)$^{11}$B reaction where much effort has been spent
recently \cite{Para91,Boyd92,Gu95,Miz00}.
Typical temperatures are around $T_9 \approx 1$
\cite{Miz00,Rau94,Boyd01} with $T_9$ being the
temperature in billion degrees. The role of light
neutron-rich nuclei in the $r$-process, e.g.\ in type II supernovae,
was analyzed in \cite{Tera01}; here a temperature range of $0.5
\lesssim T_9 \lesssim 4$ is relevant. Because of the missing Coulomb
barrier for neutron-induced reactions, astrophysically relevant
energies for the \ling\ reaction are around $E \approx kT$. Hence, for
both scenarios the cross section has to be determined for energies
below $E \lesssim 500$\,keV. In this paper all energies are given in the
center-of-mass system.

The \bpg\ reaction leads to a hot part of the $pp$-chain as soon as
the proton capture of $^8$B is faster than the competing $\beta^+$
decay \cite{Wie89}. Then a breakout to the hot CNO cyle and to the
$rp$-process is possible with the $^9$C($\alpha$,p)$^{12}$N reaction
\cite{Wie89}. The \bpg\ reaction is especially relevant in
low-metallicity stars with high masses where such a proton-rich
reaction chain can be faster than the
triple-$\alpha$ process \cite{Wie89,Ful86}, and furthermore the
reaction may become important under nova conditions
\cite{Boff93}. The typical temperature range in both astrophysical
scenarios is around several times $10^8$\,K which corresponds to
energies of the Gamow window around 50\,keV $\lesssim E \lesssim$
300\,keV. 

There are many common properties of the \ling\ and \bpg\ mirror
reactions in both experimental and theoretical point of view. Because
of the unstable $^8$Li and $^8$B nuclei, direct experiments are
extremely difficult at astrophysically relevant energies below
500\,keV. However, indirect experiments have been performed
successfully. A stringent limit for the \ling\ capture cross section
has been derived from the Coulomb breakup reaction
$^{208}$Pb($^9$Li,$^8$Li+$n$)$^{208}$Pb at MSU \cite{Zech98,Koba03}.
The astrophysical $S$-factor at zero energy for the \bpg\ reaction
(usually referred to as $S_{18}$) has been derived from the asymptotic
normalization coefficient (ANC) method using transfer reactions. The
$^2$H($^8$B,$^9$C)n reaction was measured at RIKEN \cite{Beau01}, and
one-proton removal reactions on carbon, aluminum, tin, and lead
targets were used at Texas A\&M university \cite{Tra02}. 

From theoretical point of view, the astrophysical reaction rate of
both reactions is dominated by direct (non-resonant) $E1$ transitions
from incoming $s$-waves to bound $p$-waves. However, because of the
larger Q-value of the \ling\ reaction ($Q = 4064$\,keV) and
because of the missing Coulomb repulsion the \ling\ reaction is not
purely peripheral as expected for the \bpg\ reaction with its small $Q
= 1296$\,keV.

\begin{figure}[hbt]
\includegraphics[ bb = 115 75 425 305, width = 75 mm, clip]{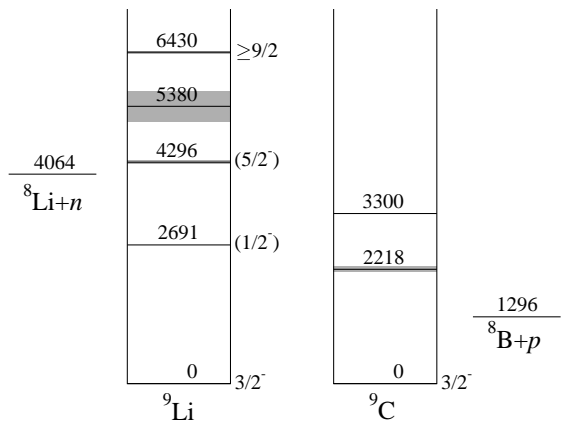}
\caption{
\label{fig:a9level} 
Level scheme of the mirror nuclei $^9$Li and $^9$C
\cite{Ajz88,NNDC,Til01}. Spin and parity of the ($5/2^-$) state at $E
= 4296$\,keV in $^9$Li are taken from theory \cite{Hees84,Rau94}.
The widths of broad levels are indicated by gray shadings.
}
\end{figure}

This paper is restricted to an analysis of the $s$-wave capture
to the ground states of $^9$Li and $^9$C. The total reaction rate for
both reactions is slightly enhanced by resonant contributions, by
$p$-wave and $d$-wave capture, and by the transition to the first
excited state in $^9$Li in the case of the \ling\ reaction. The level
scheme of the mirror nuclei $^9$Li and $^9$C (combined from
\cite{Ajz88,NNDC,Til01}) is shown in Fig.~\ref{fig:a9level}.

Various models have been used to predict the \ling\ and \bpg\ reaction
cross sections. However, practically all predictions overestimated the
experimentally determined values for both reactions
\cite{Rau94,Bert99,Desc99,Mal89,Mao91}. Especially for the \ling\
reaction, the predictions vary between a factor of 3 and up to a
factor of 50 higher than the present upper limit
\cite{Rau94,Bert99,Desc99,Mal89,Mao91} (see also Table I in
\cite{Koba03}). It is the aim of the present work to analyze the
peculiarities of the \ling\ and \bpg\ reactions at low energies.

\section{\label{sec:model}Direct Capture Model and Results}
The cross section for direct capture $\sigma^{\rm{DC}}$ is
proportional to the spectroscopic factor $C^2\,S$ and to the square of  
the overlap integral of the scattering wave function
$\chi_{\rm{scatt}}$, the electric dipole operator $O^{\rm{E1}}$, and
the bound state wave function $u_{\rm{bound}}$:
\begin{equation}
\sigma^{\rm{DC}} \sim C^2\,S \,\, 
\left| \int \chi_{\rm{scatt}}(r) \, O^{E1} \, u_{\rm{bound}}(r) \, dr \right|^2
\label{eq:dc}
\end{equation}
The full formalism can be found e.g.\ in \cite{Mohr94}. The relation
between this simple two-body model and microscopic models has been
recently studied in \cite{Esch01,Esch02}.

\begin{figure}[hbt]
\includegraphics[ bb = 150 60 490 310, width = 75 mm, clip]{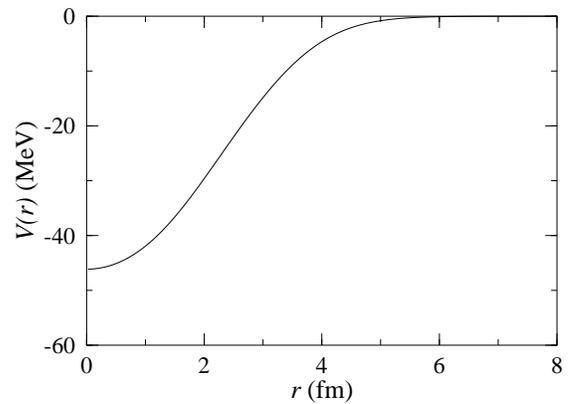}
\caption{
\label{fig:potential} 
Folding potential for the interaction $^8$Li$-n$ and $^8$B$-p$ (with
$\lambda = 1$).
}
\end{figure}

The essential ingredients are the potentials which are needed to
calculate the wave functions $\chi_{\rm{scatt}}$ and
$u_{\rm{bound}}$. In the following a real folding potential $V_F(r)$
is used
which is calculated from an approximate density for the $A = 8$ nuclei
(taken as the weighted average of the measured charge density
distributions of $^7$Li and $^9$Be \cite{Vri87}) and an effective
nucleon-nucleon interaction of $M3Y$ type \cite{Kob84}. The imaginary
part of the potential can be neglected at low energies. The resulting
potential is adjusted by a strength parameter $\lambda$ which has been
found close to unity in many cases:
\begin{equation}
V(r) = \lambda \, V_F(r) \quad .
\label{eq:lambda}
\end{equation}
For mirror reactions, it is usually accepted that the potentials
$V(r)$  and the spectroscopic factors $C^2\,S$ are very similar.
The folding potential (with $\lambda = 1$) is shown in
Fig.~\ref{fig:potential}. The volume integral per interacting nucleon
pair is $J = -616.57$\,MeV\,fm$^3$, and the root-mean-square radius
is $r_{\rm{rms}} = 3.0114$\,fm.

As usual, the parameter $\lambda$ for the bound state wave function is
adjusted to the binding energies of a 
$1p_{3/2}$ neutron (proton) in the $^9$Li ($^9$C) residual nuclei. The
resulting values are \lbo\ $ = 1.065$ (1.045) for $^9$Li
($^9$C). The deviation between both values for
\lbo\ is very small; thus the above assumption of similar
parameters for mirror nuclei is confirmed. 

For scattering waves, the potential strength parameter $\lambda$ can
be adjusted to reproduce experimental phase shifts. For $s$-waves,
which are relevant 
for the \ling\ and \bpg\ reactions, an adjustment to thermal neutron
scattering lengths is also possible (and should be preferred because
of the dominance of $s$-waves at thermal energies). Successful
examples of this procedure can be found in \cite{Beer96,Mohr98}.
Unfortunately, for the \ling\ and \bpg\ reactions phase shifts or
neutron scattering lengths are not available from experiment, and
therefore no experimental restriction from experimental scattering
data exists for \lsc .

As soon as the potential parameters \lbo\ and \lsc\ are fixed, the
capture cross sections can be calculated from Eq.~(\ref{eq:dc}); the
model contains no further adjustable parameters.
A Saxon-Woods potential can also be used, and similar results will be
obtained. But because of the larger number of adjustable parameters
the conclusions cannot be drawn as clearly as in the case of the
folding potential with the only adjustable parameter $\lambda$.

For the spectroscopic factors of the ground states of $^9$Li = $^8$Li
$\otimes\ n$ and $^9$C = $^8$B $\otimes\ p$ I use $C^2\,S = 1.0$ close
to the calculated values of $C^2\,S = 0.81 - 0.97$
\cite{Beau01,Bert99}. But also larger values up to $C^2\,S = 2.5$ have
been obtained \cite{Wie89}; this large value was not used in the
present work. It has been shown further \cite{Beau01} that the
dominating contribution to $C^2\,S$ comes from the nucleon transfer to
the $1p_{3/2}$ orbit whereas the contribution of the $1p_{1/2}$
orbit remains below 5\,\%.

\begin{figure}[hbt]
\includegraphics[ bb = 150 60 490 310, width = 75 mm, clip]{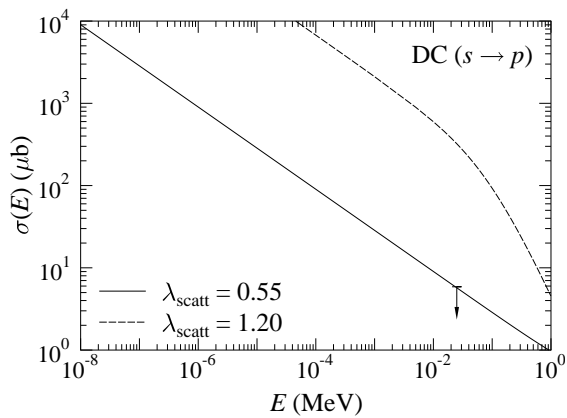}
\caption{
\label{fig:li8n} 
Direct capture cross section $\sigma^{\rm{DC}}(E)$ for the \ling\
reaction with $C^2\,S = 1.0$. The full line is obtained using \lsc\ $=
0.55$; it shows the usual $1/v$ behavior. Significant differences from
the $1/v$ behavior are found for \lsc\ $=1.20$ (dashed line). The
arrow shows the experimental upper limit from \protect\cite{Koba03}.
Discussion see Sect.~\ref{sec:disc}.
}
\end{figure}

\begin{figure}[hbt]
\includegraphics[ bb = 150 60 490 310, width = 75 mm, clip]{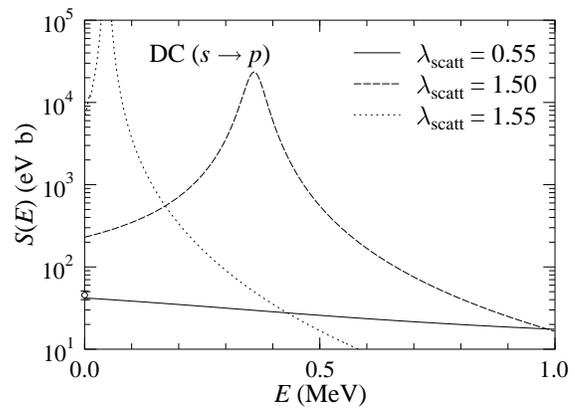}
\caption{
\label{fig:b8p} 
Astrophysical $S$-factor $S_{18}(E)$ for the \bpg\ reaction, derived from the
direct capture cross section with $C^2\,S = 1.0$. The experimental
point at $E = 0$ is taken from \protect\cite{Beau01,Tra02}. The full line is
obtained using \lsc\ $= 0.55$; as expected, the $S$-factor is almost
constant. However, resonances are obtained for \lsc\ $=1.50$ (dashed
line) and \lsc\ $= 1.55$ (dotted line). Further discussion see
Sect.~\ref{sec:disc}. 
}
\end{figure}

\begin{figure}[hbt]
\includegraphics[ bb = 145 60 490 530, width = 75 mm, clip]{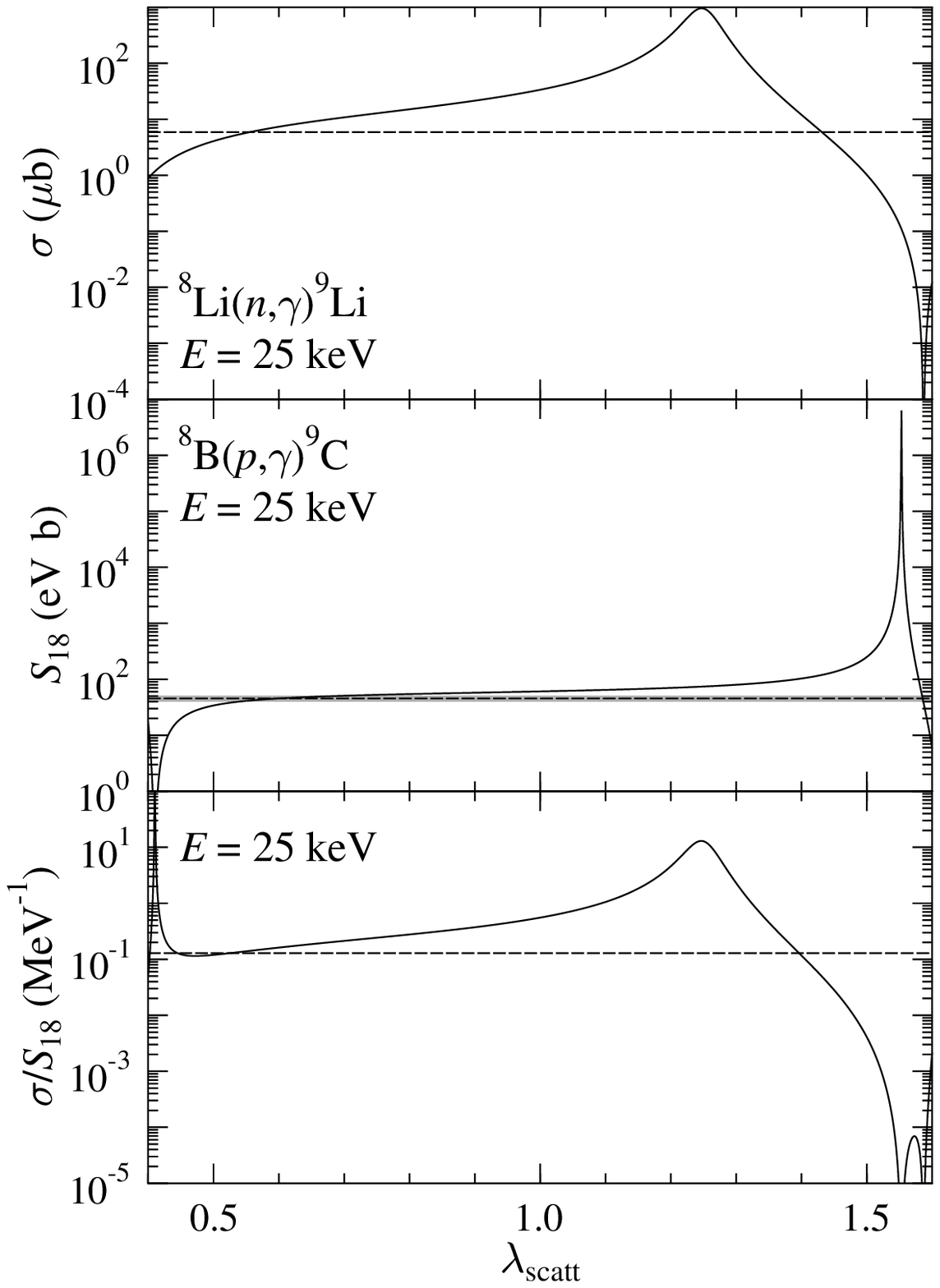}
\caption{
\label{fig:lambda} 
Direct capture cross section $\sigma^{\rm{DC}}$ for the \ling\
reaction (upper diagram), $S_{18}$ for the \bpg\ reaction (middle),
and the ratio $r = \sigma/S_{18}$ (lower), in dependence on the
potential strength parameter \lsc . All values have been 
calculated at $E = 25$\,keV using $C^2\,S = 1.0$. The horizontal lines show the
experimental results: $\sigma \le 5.88$\,$\mu$b for \ling\
\cite{Koba03}, $S_{18} = 45.5 \pm 5.5$\,eV\,b for \bpg\
\cite{Beau01,Tra02}, and the ratio $r \le 0.129$\,MeV$^{-1}$.
Discussion see text.
}
\end{figure}

To fix the potential strength parameter \lsc , and to see whether it
is possible to reproduce the experimental values of both reactions
simultaneously within this simple model, the theoretical capture cross
sections of both reactions were calculated in the energy range $E \le
1$\,MeV. A spectroscopic factor
$C^2\,S = 1.0$ was used in all calculations. In Fig.~\ref{fig:li8n}
the capture cross section $\sigma(E)$ for the \ling\ reaction is shown
as a function of energy with \lsc\ $= 0.55$ and 1.20 as parameters. In
the case of the \bpg\ reaction the energy dependence of the
astrophysical $S$-factor $S_{18}(E)$ is shown in Fig.~\ref{fig:b8p}
with \lsc\ $= 0.55$, 1.50, and 1.55 as parameters.  As can be seen,
for both reactions the results depend sensitively on the choice of the
potential strength parameter \lsc\,. Therefore for both reactions the
cross section dependence on the potential strength parameter \lsc\ has
been calculated at a fixed energy $E = 25$\,keV. The interesting
result of the cross section dependence on the potential strength
parameter \lsc\ is shown in Fig.~\ref{fig:lambda}. Note that the range
of the values for \lsc\ has to be restricted to realistic values with
the Pauli-forbidden $1s$ state below the respective threshold
($\lambda_{\rm{scatt}} \gtrsim 0.5$) and the Pauli-allowed $2s$ state
far above threshold ($\lambda_{\rm{scatt}} \lesssim
1.2$). Additionally, the ratio $r =
\sigma(25\,{\rm{keV}})/S_{18}(25\,{\rm{keV}})$ is plotted in 
Fig.~\ref{fig:lambda}. If one makes the usual assumption that the
spectroscopic factors are equal in mirror reactions, this ratio does
not depend on the chosen spectroscopic factor.

The results shown in Figs.~\ref{fig:li8n}, \ref{fig:b8p}, and
\ref{fig:lambda} have quite surprising features. It is difficult to
fit the experimental results $\sigma(25\,{\rm{keV}}) \le 5.88\,\mu$b
for the \ling\ reaction (derived from \cite{Koba03} using a standard
$1/v$ energy dependence) and $S_{18}(25\,{\rm{keV}}) \approx S_{18}(0)
= 45.5 \pm 5.5$\,eV\,b for the \bpg\ reaction (weighted average from
\cite{Beau01,Tra02}). Only in a narrow range of
$\lambda_{\rm{scatt}} \approx 0.55$ both experimental results are
reproduced simultaneously. Higher values of $\lambda$ (closer to the
expected $\lambda \approx 1$) lead to a significant overestimation of
both cross sections. From the calculated ratio $r$ in
Fig.~\ref{fig:lambda}, again the allowed range of \lsc\ is very narrow
and around \lsc\ $\approx 0.55$. Larger values of \lsc\ do not
reproduce the ratio $r$ because of the different sensitivities of the
\ling\ and \bpg\ reaction cross sections on the potential
strength. Hence it is possible to determine \lsc\ from experimental
capture data for the \ling\ and \bpg\ reactions.

A more detailed view on both \ling\ and \bpg\ reactions follows. In
Figs.~\ref{fig:li8n} and \ref{fig:b8p} the full line represents \lsc\
$= 0.55$ which was derived from the ratio of both
reactions. Additionally, the dashed lines show calculations with \lsc\
values which lead to extreme cross sections (see
Fig.~\ref{fig:lambda}).

In the case of the \ling\ reaction, the cross
section is very sensitive to the chosen value of $\lambda$. Increasing
\lsc\ from 0.55 to 0.75 leads to an increase of the cross section by
more than a factor of two, and decreasing \lsc\ from 0.55 to 0.50
reduces the cross section by about 30\,\%. With \lsc\ $= 0.55$ an
energy dependence of the cross section proportional to $1/v$ is
observed, whereas with \lsc\ $= 1.20$ a clear deviation from the usual
$1/v$ behavior can be seen in Fig.~\ref{fig:li8n}.

The \bpg\ reaction is
mainly peripheral and does not depend sensitively on the chosen
potential strength. Changing \lsc\ from 0.55 to 0.75 (0.50) increases
(decreases) the $S$-factor by 28\,\% (17\,\%). The relatively weak
dependence on 
the potential strength parameter \lsc\ can also be seen from
Fig.~\ref{fig:lambda} where $S_{18}$ changes only by roughly a factor
of two from \lsc\ $= 0.60$ to \lsc\ $= 1.4$. Furthermore, the
$S$-factor of this
reaction depends only weakly on the chosen energy; for \lsc\ $=
0.55$ one finds that $S_{18}(E=0)$ is roughly 2\,\% larger than the
quoted $S_{18}(25\,{\rm{keV}})$.

When \lsc\, is increased, in both reactions resonances are observed (see
Fig.~\ref{fig:lambda}). In the case of the \ling\ reaction the
parameter \lsc\ $= 1.25$ leads to a resonance at $E = 25$\,keV. In the
\bpg\ reaction a resonance appears at $E = 360$\,keV with a width of
$\Gamma = 50$\,keV using \lsc\ $= 1.50$; for \lsc\ $= 1.55$ this
resonance is shifted to lower energies, and the width is much smaller
(dotted line in Fig.~\ref{fig:b8p}). These resonances will be
interpreted in the following Sect.~\ref{sec:disc}.

\section{Discussion}
\label{sec:disc}
The results of the calculations now can be summarized:
\\%
($i$) There is a possibility to fit the experimental data of both
\ling\ and \bpg\ reactions within this simple model simultaneously when a
potential strength parameter \lsc\ $\approx 0.55$ and a spectroscopic
factor $C^2\,S \approx 1$ are used. Any other combinations lead to
discrepancies with the experimental results. \lsc\ is determined from
the ratio $r$ without ambiguity (independent of the spectroscopic
factors). And for much smaller or much larger values of $C^2\,S$ a
simultaneous description of both reactions is not possible.
\\%
($ii$) A surprisingly large difference for the potential parameters
\lbo\ $\approx 1$ and \lsc\ $\approx 0.55$ is
found. This might be an indication that the simple $M3Y$ interaction
fails to describe systems with extreme neutron-to-proton ratio $N/Z$;
here $N/Z = 2.0$ for $^9$Li and $N/Z = 0.5$ for $^9$C. It is
interesting to note that a calculation using a variant of the $M3Y$
interaction including a spin-orbit and a tensor part is able to
predict $S_{18}(0) = 53$\,eV\,b for the \bpg\ reaction \cite{Timo93}
which is close to the experimental results
\cite{Beau01,Tra02}. Unfortunately, there is no prediction for the
\ling\ reaction in \cite{Timo93}.
\\%
($iii$) Using \lsc\ $\gtrsim 1$ leads to a so-called ``potential
resonance''. This phenomenon has been discussed in detail in
\cite{Mohr98}, and resonances have been described successfully within
a potential model e.g.\ in \cite{Mohr94,Mohr97}. 
For $^9$Li and $^9$C the $2s$ orbit is shifted to lower energies by an
increased \lsc , and the low-energy tail of this resonance influences
the cross sections at 25\,keV significantly. For \lsc\ $\approx 1.25$
(\lsc\ $\approx 1.55$) the $2s$ resonance appears at energies around
25\,keV in the 
\ling\ (\bpg ) reaction (see Figs.~\ref{fig:li8n}, \ref{fig:b8p}, and
\ref{fig:lambda}). Probably this resonant behavior is the reason  
why most of the previous calculations using standard potentials failed
to predict the experimental data for the \ling\ and \bpg\ reactions
correctly. 

Many nuclei in the $1p$ shell with $N \approx Z$ show such
low-lying $s$-wave resonances with a large reduced width corresponding
to the $2s$ level. One example is the $1/2^+$ state at $E_x =
2365$\,keV in $^{13}$N which appears as a resonance in the
$^{12}$C($p$,$\gamma$)$^{13}$N reaction at $E = 421$\,keV \cite{Rol73}. This
resonance (and its mirror state in $^{13}$C which is located below the
$^{12}$C$-n$ threshold; this state plays an important role in the
$^{12}$C($n$,$\gamma$)$^{13}$C reaction \cite{Men95,Kik98}) can be
described within the present model using $\lambda \approx 1$
\cite{Mohr97}. Another example is the $1^-$ resonance at $E_x =
5173$\,keV in $^{14}$O which appears as a resonance in the
$^{13}$N($p$,$\gamma$)$^{14}$O reaction at $E = 545$\,keV \cite{Dec91}. 

The experimental data \cite{Koba03,Beau01,Tra02} indicate for the \ling\
and \bpg\ reactions that there are no low-lying $s$-wave
resonances close above the threshold.
What is the difference between the non-resonant \ling\ and
\bpg\ reactions and the resonant $^{12}$C($p$,$\gamma$)$^{13}$N and
$^{13}$N($p$,$\gamma$)$^{14}$O reactions? As discussed in the
following, the non-resonant behavior of \ling\ and \bpg\ can be understood from
basic shell model considerations. For simplicity, the following paragraph
discusses the \bpg\ reaction and properties of $^9$C. Similar
conclusions are reached for \ling\ and $^9$Li by exchanging protons
and neutrons. 

For light nuclei with extreme neutron-to-proton ratio $N/Z$ one has to
distinguish between neutron and proton orbits. The $J^\pi = 3/2^-$
ground state of $^9$C corresponds to a neutron $1p_{3/2}$ orbit.
Two neutrons and two protons in the $1s_{1/2}$ orbits couple to an
inert $\alpha$ core with $J^\pi = 0^+$,  and four protons fill the the
$1p_{3/2}$ subshell and couple to $J^\pi = 0^+$. 
Excited states with $J^\pi =
1/2^-$ ($1/2^+$) correspond to neutron $1p_{1/2}$ ($2s_{1/2}$)
orbits. The $1/2^-$ state is probably the first excited state (see
Fig.~\ref{fig:a9level}). The $1/2^+$ state (not known experimentally)
is expected at low energies (as usual for $1p$ shell nuclei) with a
large reduced neutron width. But the proton $2s_{1/2}$ orbit in $^9$C
must be located at much higher energies because a proton pair must be
broken for a $^9$C = $^8$B$_{\rm{g.s.}}$ $\otimes$ $p$ configuration. Spin and
parity of such a proton $2s_{1/2}$ orbit in $^9$C are $J^{\pi} = 3/2^+$ and
$5/2^+$ because $J^{\pi}(^8{\rm{B_{\rm{g.s.}}}}) = 2^+$. 

Only shallow potentials with \lsc\ $\approx 0.55$ can describe such a
high-lying proton $2s_{1/2}$ orbit in a correct way. A standard
potential (with \lsc\ $\approx 1$) would shift this proton $2s_{1/2}$
orbit to lower energies leading to a $s$-wave resonance relatively
close above the threshold. Because of the low-energy tail of this
resonance, the cross sections are strongly overestimated in this case.

The low-lying $1/2^+$ state from the neutron $2s_{1/2}$ orbit cannot
contribute to the \bpg\ reaction as $s$-wave resonance because the
excitation of a 
$J^\pi = 1/2^+$ resonance in the \bpg\ reaction requires a
$d$-wave. Additionally, only a small reduced proton width is expected
for such a neutron single-particle state.

The derivation of the equation for the overlap integral (\ref{eq:dc})
makes use of Siegert's theorem \cite{Sie37}. Therefore,
Eq.~(\ref{eq:dc}) is exactly valid only if the same Hamiltonians,
i.e.\ the same potentials, are used for the calculation of the bound
state and scattering wave functions. Otherwise, the non-orthogonality
of the bound state and scattering wave functions may lead to
considerable theoretical uncertainties \cite{Laf82}.

At first view, the huge discrepancy between the bound state potential
(\lbo\ $\approx 1.05$) and the scattering potential (\lsc\ $\approx
0.55$) seems to indicate a strong violation of Siegert's theorem.
However, this is not the case for the considered $E1$ transitions from
incoming $s$-waves in the \ling\ and \bpg\ reactions.  Practically
identical wave functions for the $p_{3/2}$ bound state are derived
from ($i$) a central potential with the strength parameter \lbo , and
($ii$) a combination of central and spin-orbit potential with strength
parameters \lbocent\ and \lbols . The shape of the spin-orbit
potential $V_{\rm{LS}}(r) = $ \lbols\ $1/r \, dV_F/dr \, \vec{L} \cdot
\vec{S}$ is practically identical to the central folding potential
$V_F(r)$ because $V_F(r)$ has an almost Gaussian shape. Hence, an
increased \lbocent\ can be compensated by a decreased \lbols\ and vice
versa, leading to the same total potential and bound state wave
function. A proper combination of \lbocent\ and \lbols\ has to be
chosen to repoduce the binding energies.  The quoted \lbo\ $\approx
1.05$ were obtained without spin-orbit potential (\lbols\ $= 0$).
Alternatively, using \lbocent\ $=$ \lsc\ $= 0.55$ leads to \lbols\ $=
-3.89$\,fm$^2$ for the $^9$Li ground state (\lbols\ $= -3.71$\,fm$^2$ for
$^9$C). The potential for the incoming $s$-wave is not affected by the
additional spin-orbit potential.  The bound state and scattering wave
functions are now orthogonal because they are calculated from the same
potential.  Therefore, the overlap integral Eq.~(\ref{eq:dc}) is exact
for the $E1$ transitions from incoming $s$-waves to bound $p$-waves if
one uses the above combination of central and spin-orbit potentials in
the entrance and exit channels.  Eq.~(\ref{eq:dc}) remains a good
aproximation if one uses only the central potential with the different
\lsc\ and \lbo\ because there are only minor deviations of the order
of 10\,\% between the different bound state wave functions; hence, the
orthogonality of the wave functions remains approximately fulfilled.
For simplicity the calculations in this work were performed without
spin-orbit potentials.

A possible core polarization (sometimes also called semi-direct
capture) may lead to asymmetries in the \ling\ and \bpg\ mirror
reactions. The core polarization can be taken into account by
modifications of the $E1$ operator in the nuclear interior. Following
the formalism of \cite{Die77} one finds that the modification of the
$E1$ operator leads to a negligible change of the cross section in the
case of the \bpg\ reaction because of its extremely peripheral
character. But also in the case of the \ling\ reaction the modified
$E1$ operator changes the cross section by significantly less than
10\,\%. Therefore, the modification of the $E1$ operator by
semi-direct capture was neglected in this work.

\section{Conclusions}
\label{sec:conc}
A consistent description of the capture cross sections of the \ling\
and \bpg\ reactions has been obtained using the direct capture
model. However, a surprisingly strong difference between the potential
parameters \lsc\ for the scattering wave function and \lbo\ for the
bound state wave function was found. The small value of \lsc\ $\approx
0.55$ is well-defined from the ratio of the cross sections of both
reactions and leads to a shallow potential. The value of \lbo\
$\approx 1.05$ is derived from the 
binding energies of the $^9$Li and $^9$C ground states and is close to
the usual values $\lambda \approx 1$. The strong
difference between \lsc\ and \lbo\ indicates limitations
of the simple $M3Y$ interaction for nuclei with extreme
neutron-to-proton ratios $N/Z$. On the other hand, the parameters
\lsc\ and \lbo\ are practically equal for the \ling\ and \bpg\ reactions,
as is expected for mirror reactions. The sensitivity of the \ling\
cross section and, to less extent, the \bpg\ $S$-factor to the
potential strength 
parameter \lsc\ is strong which can be explained by resonances in the
potential model. Consequently, the choice of potential parameters for
direct capture calculations has to be done very carefully. This
problem is not particular for the \ling\ and \bpg\ reactions, but has
to be taken into account in any direct capture calculation. The big
discrepancies between previous predictions and recently obtained
experimental results for the \ling\ and \bpg\ reactions are probably a
consequence of the neglect of these potential resonances.

\begin{acknowledgments}
I thank J.\ Escher, H.\ Oberhummer, G.\ Staudt, N.\ Timofeyuk, and A.\
Zilges for encouraging discussions.
\end{acknowledgments}

\end{document}